\documentclass[fleqn,usenatbib]{mnras}

\usepackage{newtxtext,newtxmath}

\usepackage[T1]{fontenc}

\DeclareRobustCommand{\VAN}[3]{#2}
\let\VANthebibliography\thebibliography
\def\thebibliography{\DeclareRobustCommand{\VAN}[3]{##3}\VANthebibliography}

\usepackage{graphicx}	
\usepackage{amsmath}	
\usepackage{ulem}

\title[Testing gravity with peculiar velocities]{Testing modified gravity scenarios with direct peculiar velocities}

\author[Stuart Lyall et al.]{
Stuart Lyall$^{1}$\thanks{E-mail: slyall@swin.edu.au},
Chris Blake$^{1}$,
Ryan Turner$^{1}$,
Rossana Ruggeri$^{2,1}$,
Hans Winther$^{3}$
\\
$^{1}$ Centre for Astrophysics \& Supercomputing, Swinburne
  University of Technology, P.O.\ Box 218, Hawthorn, VIC 3122,
  Australia \\
$^{2}$ School of Mathematics and Physics, University of Queensland, Brisbane, QLD 4072, Australia \\
$^{3}$ Institute of Theoretical Astrophysics, University of Oslo, 0315 Oslo, Norway
}

\date{Accepted XXX. Received YYY; in original form ZZZ}

\pubyear{2015}

\begin{document}
\label{firstpage}
\pagerange{\pageref{firstpage}--\pageref{lastpage}}
\maketitle

\begin{abstract}
The theoretical basis of dark energy remains unknown and could signify a need to modify the laws of gravity on cosmological scales.  In this study we investigate how the clustering and motions of galaxies can be used as probes of modified gravity theories, using galaxy and direct peculiar velocity auto- and cross-correlation functions.  We measure and fit these correlation functions in simulations of $\Lambda$CDM, DGP, and $f(R)$ cosmologies and, by extracting the characteristic parameters of each model, we show that these theories can be distinguished from General Relativity using these measurements.  We present forecasts showing that with sufficiently large data samples, this analysis technique is a competitive probe that can help place limits on allowed deviations from GR.  For example, a peculiar velocity survey reaching to $z=0.5$ with $20\%$ distance accuracy would constrain model parameters to 3-$\sigma$ confidence limits $\log_{10}|f_{R0}| < -6.45$ for $f(R)$ gravity and $r_c > 2.88 \, c/H_0$ for nDGP, assuming a fiducial GR model.
\end{abstract}

\begin{keywords}
dark energy -- large-scale structure of Universe -- cosmology: theory
\end{keywords}



\section{Introduction}

A wide range of cosmological observations cannot currently be explained without the inclusion of Dark Energy \citep{2013PhR...530...87W}. The current accepted model of the universe on cosmic scales is the $\Lambda$CDM model, where $\Lambda$ refers to dark energy in the form of a cosmological constant and CDM refers to cold dark matter, in the Friedmann-Robertson-Walker metric of General Relativity. The existence of Dark Energy is evident, but not well understood. A more natural explanation of spacetime expansion arising from a fundamental theory of gravity could be aesthetically preferable, as mathematically elegant formalisms have led to many progressions in scientific theory.

Many theories of modified gravity have been proposed to explain the expansion of spacetime \citep[for reviews see][]{2012PhR...513....1C,2016ARNPS..66...95J,2018LRR....22....1I,2019ARA&A..57..335F,2020Univ....6...20A}. Although none have currently distinguished themselves by better explaining observations than the standard $\Lambda$CDM model, upcoming cosmological data will allow tests at much higher levels of precision.

The observed redshift of galaxies contains components from the overall cosmic expansion, as well as from the galaxy's motion through its local space known as peculiar velocity. In order to measure the peculiar velocity, an independent distance measure must be used. This provides the cosmic expansion component of the redshift, which when subtracted from the observed redshift will leave the redshift induced from the peculiar velocity component along the line of sight. This is known as a direct peculiar velocity measurement, and has currently been performed for many thousands of local galaxies \citep[e.g.][]{2008AJ....135.1738M,2014MNRAS.445.2677S,2016AJ....152...50T}.

In linear theory, the peculiar velocity of galaxies can be expected to trace gravitational attraction towards higher density clusters \citep{1995PhR...261..271S}. Peculiar velocities of galaxies may then be used as an observable that is sensitive to proposed models that alter the force of gravity on the largest scales \citep[e.g.][]{2010MNRAS.407.2328F,2014MNRAS.445.4267K}. The rate at which matter clumps together due to this process is described by a parameter known as the growth rate of structure, the cosmic dynamics of which depend on the model of gravity. This will become the basis of our probe which uses the information from peculiar velocities to constrain gravitational physics.  In particular, we will use the correlations among and between galaxy velocities and positions as an observational probe \citep[e.g.][]{1989ApJ...344....1G,2017MNRAS.470..445N,2017MNRAS.471..839A,2021MNRAS.502.2087T}.

To demonstrate and explore the potential of the peculiar velocity probe to distinguish between modified gravity scenarios, we analyse N-body simulations which use the popular modified gravity models $f(R)$ and DGP (Dvali-Gabadadze-Poratti), as well as normal General Relativity (GR). These models are commonly analysed modified gravity theories that seek to explain cosmic expansion.

$f(R)$ theory seeks to generalise General Relativity by expressing the field equations in terms of a general function that depends on the Ricci scalar \citep[e.g.][]{2004PhRvD..70d3528C,2007PhRvD..76f4004H,2010RvMP...82..451S}. It reduces to General Relativity in low spacetime curvature environments. One common form of $f(R)$ gravity we consider in this study is the Hu-Sawicki model \citep{2007PhRvD..76f4004H}, which is parameterised by an amplitude $|f_{R0}|$, where a value of zero recovers GR. Recent constraints from cosmological analyses approximately yield $\log_{10}|f_{R0}| \lesssim -5$ \citep[e.g.][]{2012PhRvD..85l4038L,2015PhRvD..92d4009C,2016PhRvL.117e1101L}. Solar system and astrophysical tests on smaller scales are more constraining, producing $\log_{10}|f_{R0}| \lesssim -6$ \citep[e.g.][]{2007PhRvD..76f4004H,2013ApJ...779...39J,2014arXiv1409.3708S,2020PhRvD.102j4060D}.

DGP gravity suggests the four dimensional spacetime of this universe is a manifold set within a higher five dimensional space \citep{2000PhLB..485..208D}. The curvature of this space will provide an additional term to the field equations of gravity that decreases its strength on cosmic scales. This model has the free parameter of the crossover scale $r_c$, which roughly describes the scale needed to observe significant deviations from GR. We will specifically consider the normal branch DGP model \citep{2003JCAP...11..014S,2004PhRvD..70j1501L} for which the background evolution is the same as in $\Lambda$CDM. The Cosmic Microwave Background (CMB) measurements provide the constraint $r_c>3.5c/H_0$ to 95\% confidence \citep{2009PhRvD..80f3536L}.

Peculiar velocities are becoming an increasingly important probe of cosmic dynamics in precision cosmology. As such, a number of surveys have been conducted to measure peculiar velocities, and the power of these datasets is rapidly growing. The 6-degree Field Galaxy Survey (6dFGS) measured peculiar velocities from 8885 galaxies up to a redshift of $z=0.055$ using the fundamental plane to determine distances \citep{2014MNRAS.445.2677S}. Cosmicflows-3 \citep{2016AJ....152...50T} combines the 6dFGS data with other surveys to list 17669 measured peculiar velocities, using in addition the Tully-Fisher relation. These samples will be greatly expanded by the Dark Energy Spectroscopic Instrument, which is currently obtaining a large set of peculiar velocity data that reaches $z=0.1$ as part of its Bright Galaxy Survey \citep{2016arXiv161100036D}, and the 4-metre Multi-Object Spectrograph Telescope (4MOST) Hemisphere Survey.

Several previous analyses have used cosmological simulations to consider the impact of modified gravity scenarios on peculiar velocity statistics.  \cite{2014PhRvL.112v1102H} demonstrated that the low-order 
moments of the galaxy pairwise velocity distribution exhibit significant deviations from general relativity in $f(R)$ and Galileon gravity models. \cite{2016MNRAS.462...75S} simulated the magnitude of bulk 
velocity flows in both $\Lambda$CDM and $f(R)$ cosmologies, inferring limits on the characteristic amplitude $f_{R0}$ from bulk flow measurements.  Most recently, \cite{2022arXiv220914386N} presented the first constrained simulations of the density and velocity fields of the local Universe in DGP and $f(R)$ models.  The redshift-space clustering of halos in $f(R)$ and DGP cosmologies has been studied in detail by a number of authors including \cite{2012MNRAS.425.2128J,2016PhRvD..94h4022B,2019MNRAS.485.2194H,2021PhRvD.103j3524G,2022arXiv220801345F}.

We structure our paper as follows.  In Section \ref{sectheory} we overview the modified gravity scenarios and cosmological theory we consider in our study.  In Section \ref{secsim} we introduce the modified gravity simulations we use to test our analysis, and in Section \ref{secmethod} we lay out our analysis process to extract cosmic parameters.  In Section \ref{secresults} we present the results of applying our analysis to the simulations we use. In Section \ref{secforecast} we use Fisher matrices to forecast the constraining power of this analysis if applied to upcoming future galaxy surveys, and we conclude in Section \ref{secconc}.

\section{Theory}
\label{sectheory}

\subsection{Growth rate of structure}

The growth rate of cosmic structure is a powerful observable for analysing the evolution of large-scale structure and differentiating between cosmological models. The growth rate of structure $f$ is defined by,
\begin{equation}
f=\frac{d\ln(\delta_m)}{d\ln(a)} ,
\label{eq_growth}
\end{equation}
where $\delta_m$ is the overdensity of a perturbation and $a$ is the cosmic scale factor.  The growth rate measures the logarithmic derivative of the clustering amplitude with scale factor, as the universe evolves.

The dynamics of $f$ and the clustering of matter can be determined from linear perturbations of the density field.  The linear matter perturbation equation can be formed from the equation of motion of a fluid,
\begin{equation}
\frac{\partial\underline{v}}{\partial t}+\left(\underline{v}\cdot\nabla\right)\underline{v}=-\nabla\Phi ,
\label{eq_fluid}
\end{equation}
where $\underline{v}$ represents the flow velocity field of matter, and $\Phi$ is the gravitational potential. The continuity equation provides a link between the flow divergence and density:
\begin{equation}
\frac{\partial\rho}{\partial t}+\nabla\cdot(\rho\underline{v})=0 .
\label{eq_continuity}
\end{equation}
Finally, the Poisson equation in standard gravity connects the second derivative of the potential and density:
\begin{equation}
\nabla^2\Phi=4\pi G\rho ,
\label{eq_poisson}
\end{equation}
where $G$ is Newton's universal gravitational constant.

After substituting Eq.\ref{eq_continuity} and Eq.\ref{eq_poisson} in Eq.\ref{eq_fluid} and reducing these equations to their linear form, we find
\begin{equation}
\ddot{\delta}_m+2H\dot{\delta}_m-4\pi G\overline{\rho}\delta_m=0 ,
\label{eq_linpert}
\end{equation}
where $H=\dot{a}/a$ is the Hubble parameter and $\overline{\rho}$ is the mean cosmic density. From Eq.\ref{eq_growth} we can also find the relation,
\begin{equation}
\dot{\delta}_m=Hf\delta_m .
\label{eq_oddot2f}
\end{equation}
This allows the linear perturbation equation to be reparameterised in terms of the growth rate of structure as follows:
\begin{equation}
\frac{df}{da}=\frac{3G_{\rm eff} H_0^2 \Omega_{m0}}{2 a^4 H^2}-\left(\frac{1}{H}\frac{dH}{da}+\frac{2}{a}\right)f-\frac{f^2}{a} .
\label{eq_linpertf}
\end{equation}
We have replaced the gravitational constant in Eq.\ref{eq_linpertf} by $G_{\rm eff}$, or the effective gravitational constant, which is a parameterisation which can be used to describe the diverging effects of a modified gravity model. In this equation $\Omega_{m0}$ represents the current density of matter normalised by the critical density $\rho_c = 3H^2/8\pi G$.  Eq.\ref{eq_linpertf} then forms the equation of motion for $f$ \citep[e.g.][]{2005PhRvD..72d3529L}.  As we shall see, a key characteristic differentiating models is whether $G_{\rm eff}$ (and hence the growth rate) depends on scale.

The Friedmann equations for cosmic evolution allow $H$ to be parameterised in terms of $a$ for a given cosmology. The current $\Lambda$CDM model leads to the form,
\begin{equation}
H^2=H_0^2\left(\Omega_{R0}a^{-4}+\Omega_{m0}a^{-3}+\Omega_{k0}a^{-2}+\Omega_{\Lambda0}\right) ,
\label{eq_friedmann}
\end{equation}
with $\Omega_{R0}$ representing the current radiation density, $\Omega_{\Lambda0}$ representing the current dark energy density, and $\Omega_{k0}$ representing the magnitude of effects due to spatial curvature.

The Friedmann equations originate from the field equations that describe the relationship between the curvature of space and its matter-energy content, which lies at the heart of how gravity is described. This is the same root relationship which also contributes to Eq.\ref{eq_linpertf}. Modified gravity theories will necessarily have different field equations, and will then also have altered Friedmann equations.

\subsection{General Relativity}

General Relativity is the current best experimentally-verified theory of gravity. The link between matter and spacetime curvature used by the $\Lambda$CDM model is mathematically expressed by Einstein's field equations:
\begin{equation}
R_{\alpha\beta}-\tfrac{1}{2}Rg_{\alpha\beta}=\tfrac{8\pi G}{c^4}T_{\alpha\beta}-\Lambda g_{\alpha\beta} ,
\label{eq_EFE}
\end{equation}
where $R_{\alpha\beta}$ is the Ricci curvature tensor, $R$ is the Ricci curvature scalar, $g_{\alpha\beta}$ is the spacetime metric, $T_{\alpha\beta}$ is the covariant form of the energy momentum tensor, $c$ is the speed of light, and $\Lambda$ is the cosmological constant.

In the search for modified theories of gravity, it is productive to alter or generalise Eq.\ref{eq_EFE}. Due to extensive testing that has not disproven GR on solar-system scales \citep[e.g.][]{2012PhR...513....1C,2014LRR....17....4W,2018LRR....22....1I,2019ARA&A..57..335F}, any successful modified theory of gravity seeking to explain cosmic evolution should approximate GR in the small-scale limit of our current observed regimes.

It's convenient when describing modern physical laws to express them in the more compact form of a least-action principle. Einstein's field equations can be reformatted as the Einstein-Hilbert action, where we include the cosmological constant:
\begin{equation}
S=\int\frac{R-2\Lambda}{16\pi G}\sqrt{|g|}dx^4 ,
\label{eq_EHaction}
\end{equation}
where the integral is taken across all four dimensions of spacetime, and $|g|$ refers to the determinant of the metric.

\begin{figure}
    \centering
    \includegraphics[width=\columnwidth]{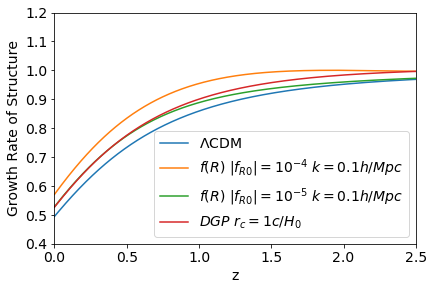}
    \caption{The evolution of the growth rate of structure in the redshift range $0 < z < 2.5$ for different gravity models. We compare the theoretically-calculated growth rates for $\Lambda$CDM, $f(R)$ models with two different values of $|f_{R0}|$ for $k = 0.1 \, h$ Mpc$^{-1}$, and a DGP model with $r_c = 1c/H_0$. These differences in growth rate can be used to distinguish between these models.}
    \label{fig_fmodels}
\end{figure}

When using Eq.\ref{eq_linpertf} to compute the growth rate in GR, we set $G_{\rm eff}=1$.  Fig.\ref{fig_fmodels} illustrates the evolution of the growth rate of structure with redshift for our fiducial $\Lambda$CDM model. For this model and all others we assume $\Omega_{m0}=0.2815$ and $\Omega_{\Lambda0}=0.7185$ (matching the fiducial cosmology of the simulations we will introduce in Section \ref{secsim}).

\subsection{\texorpdfstring{$f(R)$}{f(R)} model}
\label{sec:fr}

The $f(R)$ modified gravity model seeks to generalise gravity by adding a function of the Ricci scalar to the Einstein-Hilbert action, denoted as $f(R)$ \citep{2004PhRvD..70d3528C,2007PhRvD..76f4004H,2010RvMP...82..451S,2019JCAP...09..066M}:
\begin{equation}
S=\int\frac{f(R)}{16\pi G}\sqrt{|g|}dx^4 .
\label{eq_fRaction}
\end{equation}
This is a mathematically simple way to explore a broad range of possible extensions to GR. The dependence on the Ricci scalar means that the modified field equations will also satisfy the conventional energy conservation constraints that motivated the original field equations.

The altered field equations obtained from this action are given by,
\begin{equation}
\frac{df(R)}{dR}R_{\mu\nu}-\frac{1}{2}f(R)g_{\mu\nu}+\left(g_{\mu\nu}\nabla^\lambda\nabla_\lambda-\nabla_\mu\nabla_\nu\right)\frac{df(R)}{dR}=\frac{8\pi G}{c^4}T_{\mu\nu} .
\label{eq_fRFE}
\end{equation}
The specific function $f(R)$ analysed in our study is the Hu-Sawicki model with its $n$ parameter equal to $1$ \citep{2007PhRvD..76f4004H}:
\begin{equation}
f(R)=R-2\Lambda-f_{R0}\frac{R_0^2}{R} ,
\end{equation}
where $R_0$ refers to the average Ricci scalar today and $f_{R0}$ is a free parameter which describes the amplitude of the divergence from GR. The Hu-Sawicki model is a simple yet non-trivial modified theory that approaches GR when $R$ is sufficiently large; when there is curvature in spacetime or a gravitational potential is present. This renders the model indistinguishable from conventional GR in the presence of dense environments or gravitational potential wells, but allows deviations in cosmic voids and in large low density regions between galaxy clusters. This suppression dependence is said to screen diverging effects via the chameleon screening method \citep{2004PhRvD..69d4026K}.

\begin{figure}
    \centering
    \includegraphics[width=\columnwidth]{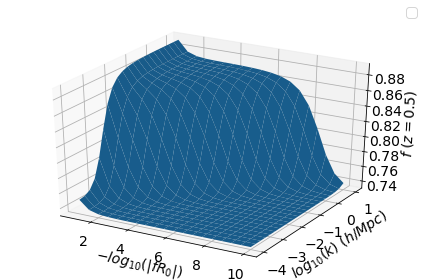}
    \caption{The growth rate of structure at redshift $z=0.5$ for the $f(R)$ model, as a function of the parameter $|f_{R0}|$ which describes the strength of the deviation from GR, and wavenumber $k$.  The scale dependence of the growth rate is evident, with ever-smaller scale modes being influenced as the $f(R)$ model becomes more deviant from General Relativity.}
    \label{fig_fvfR0vscale}
\end{figure}

The effective gravitational constant for the matter perturbation equation in $f(R)$ gravity is given by \cite{2019JCAP...09..066M}:
\begin{equation}
G_{\rm eff}(k) = \left(\frac{df(R)}{dR}\right)^{-1}\left(1+\frac{1}{\frac{a^2}{k^2}\left(\frac{d^2f(R)}{dR^2}\right)^{-1}+3}\right) ,
\label{eq_GeffR}
\end{equation}
where $k$ represents the fluctuation wavenumber being considered. Unlike General Relativity and DGP models, the growth rate of $f(R)$ is hence dependent on the scale, which can aid in distinguishing its effects. Fig.\ref{fig_fmodels} shows how different $f_{R0}$ values will alter the structure growth rate history for some examples. Fig.\ref{fig_fvfR0vscale} highlights the growth rate of structure's dependence on scale in $f(R)$ cosmologies. As can be seen, a measurable divergence from GR only occurs for structure modes above a certain scale, with a larger range of modes affected at larger $f_{R0}$ values.

\subsection{DGP model}

The Dvali-Gabadadze-Porrati (DGP) gravity model generalises General Relativity by embedding four dimensional spacetime in a five dimensional manifold \citep{2000PhLB..485..208D}. We will specifically consider the normal branch of DGP \citep{2003JCAP...11..014S,2004PhRvD..70j1501L}, sometimes referred to as nDGP, though the background expansion of space is the treated the same as $\Lambda$CDM in our study (see Sec.\ \ref{secsim}). The normal branch models are named to distinguish them from the self-accelerating class of models which have been discredited as viable theories \citep[e.g.][]{2008PhRvD..78j3509F}. The curvature of this higher-dimensional space provides another term in the Hilbert action and allows the observed gravitational forces to be altered:
\begin{equation}
S=\int\frac{R^{(5)}}{16\pi G^{(5)}}\sqrt{|g|^{(5)}}dx^5+\int\bigg(\frac{R}{16\pi G}+\mathcal{L}^{(m)}\bigg)\sqrt{|g|}dx^4 ,
\end{equation}
where $G^{(5)}$ represents the strength of the coupling between energy and spatial curvature in the 5D space, $R^{(5)}$ is the 5D Ricci tensor, $|g|^{(5)}$ represents the determinant of the metric for the 5D space, and $\mathcal{L}^{(m)}$ represents the matter action component that describes the motion of everyday matter which can be noted to only freely move in the 4D brane.

The relative strength of the additional term, compared to conventional gravity, is quantified by the crossover scale parameter $r_c$:
\begin{equation}
r_c=\frac{1}{2}\frac{G^{(5)}}{G} .
\end{equation}
The divergent effects of this gravity model will only become detectable on scales larger than this crossover scale: at shorter distances, the model becomes screened and General Relativity is reproduced.

The effective gravitational constant in Eq.\ref{eq_linpertf}, which characterises the evolution of $f$ in the case of DGP gravity, is given by \cite{2016PhRvD..94h4022B}:
\begin{equation}
G_{\rm eff}=1+\frac{1}{3 \left[ 1+2\frac{r_c}{c}\left( H+\frac{1}{3}\frac{dH}{d\ln(a)} \right) \right]} .
\end{equation}
In Fig.\ref{fig_fmodels}, the growth history of a DGP model with $r_c=1c/H_0$ is displayed.

\section{Simulations}
\label{secsim}

We conducted our study by utilising existing N-body simulations generated for standard General Relativity, the DGP model and $f(R)$ gravity.  These simulations are fully described in \cite{2015MNRAS.454.4208W}, \cite{2018MNRAS.476.3195C} and \cite{2020arXiv201105771A} and we briefly summarise the key details here.

The DGP model simulations use the normal branch of the DGP model, where we utilised simulations with two different crossover parameters $r_c = (1c/H_0, 5c/H_0)$ which we label (N1, N5).  The $f(R)$ model simulations use the Hu-Sawicki model with $n=1$, where we used simulations with three different model amplitudes $|f_{R0}| = (10^{-4}, 10^{-5}, 10^{-6})$ which we label (F4, F5, F6).  For each model cosmology, we analysed five independent simulation boxes, whose measurements could be averaged to produce more accurate results.

All of these simulations were created with the property that their background expansion is fixed to the same fiducial $\Lambda$CDM expansion history as used in the GR simulation.  Hence, the only modifications are in the growth of structure (i.e., additional fifth force).  This feature enables us to study the effects of the different gravitational physics independently of the expansion history of each simulation.

Furthermore, all simulations were generated with exactly the same initial conditions, i.e.\ the same initial flat $\Lambda$CDM power spectrum at early times, which corresponds to a GR model with cosmological parameters $\Omega_m = 0.2814$, $\Omega_b = 0.0464$, $h = 0.697$, $n_s = 0.971$ and $\sigma_8 = 0.82$ (a WMAP9 cosmology).  The simulations hence represent Universes that evolve identically until late times, where the fifth-force takes effect.  In this case the effect of cosmic variance on the ratios of power spectra will be suppressed on large scales.  However, the value of $\sigma_8$ today will vary between the simulations owing to their different growth histories, which we include in our fitting approach.

The simulations were generated in a box of side 1024 $h^{-1}$ Mpc with periodic boundary conditions and populated by $1024^3$ particles each of mass $7.8 \times 10^{10} \, h^{-1} M_\odot$. The simulations used a particle snapshot at redshift $z=0.5$, which is close to the median effective redshift of the BOSS survey \citep{2013AJ....145...10D}, which inspired the generation of the simulations. Halo-finding was performed using the RockStar code \citep{2013ApJ...762..109B}, and the halos were populated with galaxies using a halo occupation distribution which matched the projected 2-point galaxy correlation function (that is, the real-space clustering) of the BOSS CMASS sample \citep{2013MNRAS.428.1036M} in all cases.  Hence, the main distinguishing feature of the simulations is the properties of their velocity field, which is the main point of interest for our study.

\section{Methods}
\label{secmethod}

\subsection{Correlation function measurements}
\label{sec_corrmeas}

We measured the galaxy and velocity auto- and cross-correlations from the simulated catalogues.  For the purposes of this analysis we used the projected velocity along the separation vector of two galaxies, rather than the line-of-sight velocity along an axis to which observations would be restricted, and applied no additional observational velocity error, in order to achieve greater precision in our tests and specify any systematic errors as accurately as possible.  We also used real-space rather than redshift-space positions for our analysis, to allow a cleaner test of the linear component of the growth rate without specifying any non-linear Redshift Space Distortion (RSD) corrections.  However, we will forecast observational results based on a line-of-sight velocity, realistic noise and RSD in Section \ref{secforecast}.

For each simulation dataset, we randomly selected a sub-sample of $10^5$ haloes with positions and velocities. This sub-sampling allowed us to perform correlation function measurements more efficiently, without significantly increasing the measurement error, which is limited by sample variance. We created a corresponding random set for each sample, containing $10^6$ particles randomly distributed across the same space, and random velocities sampled from a multidimensional Gaussian distribution with the same velocity distribution as the data sample.

\begin{figure*}
    \centering
    \includegraphics[width=16cm]{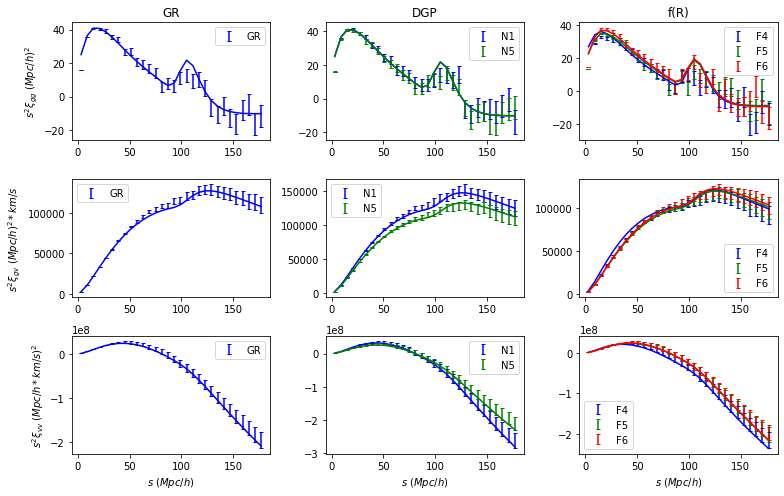}
    \caption{The averaged correlation function measurements from the $z=0.5$ snapshots of the simulations for all 6 considered cosmologies, over-plotting the best-fitting theoretical correlation function models. The panels are separated by the gravity model (left to right: General Relativity, DGP, $f(R)$), and by observational statistic (top to bottom: galaxy auto-correlation, galaxy-velocity cross-correlation, velocity auto-correlation).}
    \label{fig_corrfit}
\end{figure*}

The correlation functions were then calculated from the data and random samples via the estimators proposed by \cite{1993ApJ...412...64L} and expanded for velocity statistics by \cite{2014JCAP...05..003O} and \cite{2021MNRAS.502.2087T}:
\begin{equation}
\hat{\xi}_{\delta\delta}=\frac{N_R^2}{N_D^2}\frac{D_\delta D_\delta}{R_\delta R_\delta}-2\frac{N_R}{N_D}\frac{D_\delta R_\delta}{R_\delta R_\delta}+1 ,
\end{equation}
\begin{equation}
\hat{\xi}_{\delta v}=\frac{N_R^2}{N_D^2}\frac{D_\delta D_v}{R_\delta R_\delta}-\frac{N_R}{N_D}\frac{D_\delta R_v}{R_\delta R_\delta}-\frac{N_R}{N_D}\frac{R_\delta D_v}{R_\delta R_\delta}+\frac{R_\delta R_v}{R_\delta R_\delta} ,
\end{equation}
\begin{equation}
\hat{\xi}_{vv}=\frac{N_R^2}{N_D^2}\frac{D_vD_v}{R_\delta R_\delta}-2\frac{N_R}{N_D}\frac{D_v R_v}{R_\delta R_\delta}-\frac{D_vR_v}{R_\delta R_\delta} ,
\end{equation}
where $\xi_{\delta\delta}$ is the galaxy auto-correlation, $\xi_{\delta v}$ is the galaxy-velocity cross-correlation, $\xi_{vv}$ is the velocity auto-correlation, $N_D$ is the number of particles in the data sample, and $N_R$ is the number of particles in the random sample, which we take to be 10 times higher than $N_D$. These estimators also contain different pair counts between the data galaxy and velocity samples (denoted by $D_\delta$ and $D_v$) and the corresponding random samples (denoted by $R_\delta$ and $R_v$), where velocity pair counts are weighted by the inward velocity of each component particle projected along the separation vector to the other particle.  Given the simulations are generated with periodic boundary conditions, we measured pair counts in a periodic box, wrapping the catalogue and retaining unique pairs.  This increases the accuracy of our covariance matrix estimate.  We measured correlation functions in the separation range 0 to 180 $h^{-1}$ Mpc, split into 30 separation bins each of width 6 $h^{-1}$ Mpc.  We made measurements for the 5 realisations of each modified gravity scenario, and averaged the results. Fig.\ref{fig_corrfit} shows the correlation function measurements extracted from the simulations.

\subsection{Correlation function models}
\label{sec_corrmod}

In this section we outline our calculation of the theoretical correlation functions in each model of gravitational physics, for comparison with the measurements.  We adopt a linear-theory model which we apply to large-scale measurements as discussed in Sec.\ \ref{sec:parfit}.  The auto- and cross-correlation functions between galaxy overdensity $\delta$ and infall velocity $v$ at separation vector $\underline{r}$ are defined by the following averages over space:
\begin{equation}
\xi_{\delta\delta}(\underline{r}) = b^2 \, \langle \delta_m(\underline{x}) \, \delta_m(\underline{x}+\underline{r}) \rangle ,
\label{eq_corgg}
\end{equation}
\begin{equation}
\xi_{\delta v}(\underline{r}) = b \, \langle \delta_m(\underline{x}) \, \left[ \underline{v}(\underline{x}+\underline{r})\cdot(-\underline{r}) \right] \rangle ,
\label{eq_corgv}
\end{equation}
\begin{equation}
\xi_{vv}(\underline{r}) = \langle  \left[ \underline{v}(\underline{x})\cdot\underline{r} \right] \, \left[ \underline{v}(\underline{x}+\underline{r})\cdot(-\underline{r}) \right] \rangle ,
\label{eq_corvv}
\end{equation}
where $b$ is the linear galaxy bias factor, and the $\langle ... \rangle$ brackets refer to the expected outcome averaged across all $\underline{x}$.  Here we adopt a linear bias approximation for how the observable galaxy field $\delta$ relates to the total matter density field $\delta_m$, given that we are studying fluctuations on the largest scales.

It is convenient to express these correlation functions in terms of the matter power spectrum $P(k)$ as a function of wavenumber $k$, since this is the quantity predicted most directly by models of the early Universe.  This is also advantageous because different Fourier density modes evolve independently in linear theory.  The matter power spectrum is defined as:
\begin{equation}
P(\underline{k})=\langle \Tilde{\delta}_m(\underline{k}) \, \Tilde{\delta}_m^*(\underline{k}) \rangle .
\label{eq_power}
\end{equation}
For a matter field that has a curl-free velocity and satisfies the continuity equation, the Fourier transform of the velocity field can be derived as \citep[e.g.][]{2017MNRAS.471..839A},
\begin{equation}
\underline{\Tilde{v}}(\underline{k}) = -iaHf\frac{\underline{k}}{k^2}\Tilde{\delta}_m(\underline{k}) ,
\label{eq_velf}
\end{equation}
where $i$ is the imaginary number and $\underline{k}$ is a 3D wavevector.  With appropriate substitutions of Eq.\ref{eq_power} and Eq.\ref{eq_velf}, the average of Eq.\ref{eq_corgg}-\ref{eq_corvv} over all directions can be expressed as:
\begin{equation}
\xi_{\delta\delta}(r)=\frac{b^2}{2\pi^2r} \int k \, \sin(kr) \, P(k) \, dk ,
\label{eq_corggf}
\end{equation}
\begin{equation}
\xi_{\delta v}(r)=\frac{b a H}{2\pi^2}\int f(k) \, k \, j_1(kr) \, P(k) \, dk ,
\label{eq_corgvf}
\end{equation}
\begin{equation}
\xi_{vv}(r)=\frac{a^2 H^2}{2\pi^2}\int f^2(k) \, k \, \left(j_0(kr)-\frac{2j_1(kr)}{kr}\right) \, P(k) \, dk ,
\label{eq_corvvf}
\end{equation}
where $j_n$ represents a spherical Bessel function of the first kind.

We note that the growth rate of structure is placed inside the integral over scale in Eq.\ref{eq_corggf}-\ref{eq_corvvf}.  In General Relativity the growth of structure is independent of scale in linear theory, but this is not necessarily true for all models of gravity.  For example, the growth rate of structure for $f(R)$ gravity is dependent on scale, as can be seen in Eq.\ref{eq_GeffR} and Fig.\ref{fig_fmodels}.

The correlation function models are now specified by the redshift of observation, expansion rate $H$, scale-dependent growth rate $f(k)$ and power spectrum $P(k)$ at this redshift.  As discussed in Section \ref{secsim}, the simulations investigated here all follow the same expansion history and same initial power spectrum at early times.  In this case the power spectrum $P(k)$ of the simulations at a given redshift will differ by the growth history in each specific model.  This can be expressed in terms of the growth factor $g$ which describes the scaling of the matter fluctuations $\delta_m$.  From Eq.\ref{eq_growth},
\begin{equation}
f=\frac{d\ln(g)}{d\ln(a)} .
\label{eq_growfacf}
\end{equation}
With a growth rate $f(k,a)$ determined from solving the growth differential equation, Eq.\ref{eq_growfacf} can be rearranged such that:
\begin{equation}
g(k,a) = g_0 \, \exp{\left[ -\int_{a_{\rm ini}}^a \frac{f(k,a')}{a'} \, da' \right]} ,
\label{eq_growfac}
\end{equation}
where $g_0$ is the initial value at early times $a = a_{\rm ini}$ which is matched across the simulations.  This factor cancels out when the growth factor is normalised while being applied to the generalised power spectrum equation below:
\begin{equation}
P(k,a) = P_{GR}(k,a) \, \frac{g^2(k,a)}{g^2_{GR}(a)} ,
\label{eq_powerMG}
\end{equation}
where $P_{GR}$ is the fiducial power spectrum, for which we use a non-linear model generated with the CAMB software \citep{2000ApJ...538..473L} with the same input cosmological parameters as used to generate our simulations described in Section \ref{secsim}, and $g_{GR}$ is the growth factor found for the General Relativity model. The only parameter needed to evaluate the correlation function is then the growth rate of structure. In order to calculate the growth factor, Eq.\ref{eq_linpertf} was solved numerically with the use of the Runge-Kutta method. The growth factor was then calculated using Eq.\ref{eq_growfac}.

The $f(R)$ simulation includes significant nonlinear screening effects which cause the linear growth computed in Sec.\ \ref{sec:fr} to be inaccurate. To account for this scale-dependent effect, the $f(R)$ growth factor was modified using the phenomenological screening relation \citep[e.g.][]{2017PhRvD..95f3502A},
\begin{equation}
g(k) = g_{GR} + \left[ g_{f(R)}(k)-g_{GR} \right] \, e^{-(kr_s)^2} ,
\label{eq_fRscreen}
\end{equation}
where $r_s$ is the screening scale beyond which the growth factor asymptotes to the GR growth factor.  We selected a value of $r_s=30 \, h^{-1}$ Mpc as a good fit to our simulations, noting that marginalizing over this value did not significantly change our results (see Sec.\ \ref{secfrfit} for more details).

\subsection{Analytical covariance matrix}
\label{sec_corrcov}

We generated an analytical covariance matrix for the joint galaxy and velocity correlation functions, assuming Gaussian-distributed variables.  Whilst an approximation, this should be sufficiently accurate on large scales, given that velocity statistics are dominated by low-$k$ modes.  We averaged each correlation function $\xi(r)$ over angles and in a separation bin $i$ in the range $r_1 < r < r_2$, such that
\begin{equation}
\overline{\xi}^i = \frac{\int_{r_1}^{r_2} dr \, r^2 \, \xi(r)}{\int_{r_1}^{r_2} dr \, r^2} .
\label{eqxiave}
\end{equation}
The analytical covariance between the bin-averaged correlation functions has the general form which we derive in Appendix \ref{secapp},
\begin{equation}
  {\rm Cov} \left[ \overline{\xi}^i_A , \overline{\xi}^j_B \right] = \frac{2}{V} \int \frac{dk \, k^2}{2\pi^2} {\rm Cov} \left[ P_A(k), P_B(k) \right] \, W^i_A(k) \, W^j_B(k) ,
\end{equation}
where $(A,B)$ represent correlation function types $(\delta\delta, \delta v, vv)$, $V$ is the volume of the Fourier cuboid, $P_{A,B}$ is the power spectrum associated with each correlation function, and $W^i$ represents a window function depending on the correlation function type,
\begin{equation}
W^i_{\delta \delta} = \frac{3 \left[ r_2^2 \, j_1(kr_2) - r_1^2 \, j_1(kr_1) \right]}{k \left( r_2^3 - r_1^3 \right)} ,
\end{equation}
\begin{equation}
W^i_{vv} = \frac{3 \int_{k r_1}^{k r_2} du \, u^2 \left[ j_0(u) - \frac{2 j_1(u)}{u} \right]}{k^3 \left( r_2^3 - r_1^3 \right)} ,
\end{equation}
\begin{equation}
W^i_{\delta v} = \frac{3 \int_{k r_1}^{k r_2} du \, u^2 \, j_1(u)}{k^3 \left( r_2^3 - r_1^3 \right)} ,
\end{equation}
where these expressions are derived in Appendix \ref{secapp}.  The covariance terms for the different combinations are,
\begin{equation}
  {\rm Cov} \left[ P_{\delta \delta}(k) , P_{\delta \delta}(k) \right] = \left[ P_{\delta \delta}(k) + \frac{1}{n_g} \right]^2 ,
\label{eqpggcov}
\end{equation}
\begin{equation}
  {\rm Cov} \left[ P_{\delta \delta}(k) , P_{\delta v}(k) \right] = \left[ P_{\delta \delta}(k) + \frac{1}{n_g} \right] \, P_{\delta v}(k) ,
\end{equation}
\begin{equation}
  {\rm Cov} \left[ P_{\delta \delta}(k) , P_{vv}(k) \right] = P_{\delta v}(k)^2 ,
\end{equation}
\begin{equation}
  {\rm Cov} \left[ P_{vv}(k) , P_{vv}(k) \right] = \left[ P_{vv}(k) + \frac{\sigma_v^2}{n_g} \right]^2 ,
\label{eqpvvcov}
\end{equation}
\begin{equation}
  {\rm Cov} \left[ P_{vv}(k) , P_{\delta v}(k) \right] = \left[ P_{vv}(k) + \frac{\sigma_v^2}{n_g} \right] \, P_{\delta v}(k) ,
\end{equation}
\begin{equation}
\begin{split}
  &{\rm Cov} \left[ P_{\delta v}(k) , P_{\delta v}(k) \right] = \\ &\frac{1}{2} \left( P_{\delta v}(k)^2 + \left[ P_{\delta \delta}(k) + \frac{1}{n_g} \right] \left[ P_{vv}(k) + \frac{\sigma_v^2}{n_g} \right] \right) ,
\label{eqpgvcov}
\end{split}
\end{equation}
where $n_g$ is the galaxy number density and $\sigma_v$ is the dispersion of velocities due to noise.  Assuming a GR model with linear bias $b$, the different power spectra are related to the matter power spectrum $P_m(k)$ as $P_{\delta \delta}(k) = b^2 \, P_m(k)$, $P_{vv}(k) = \frac{a^2 H^2 f^2}{k^2} P_m(k)$ and $P_{\delta v}(k) = \frac{a H f b}{k} P_m(k)$, where we emphasize again that $v$ is the projected velocity along the vector separation, rather than the line-of-sight velocity along an axis.

\begin{figure}
    \centering
    \includegraphics[width=\columnwidth]{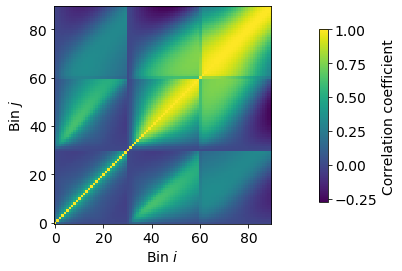}
    \caption{The theoretical covariance matrix between the three correlation functions for the GR simulation and all separation bins. The overdensity auto correlation is shown at low index value (bin 1-30), followed by the galaxy-velocity cross-correlation function (bin 31-60), and then the velocity auto-correlation is shown at high index values (bin 61-90).  The covariance matrix $C_{ij}$ is displayed as a correlation matrix $C_{ij}/\sqrt{C_{ii} \, C_{jj}}$.}
    \label{fig_covmat}
\end{figure}

We evaluated the analytical covariance for the fiducial model of each modified gravity simulation, assuming a non-linear model power spectrum generated using the simulation initial conditions.  We considered a box volume $V = (1024 \, h^{-1} {\rm Mpc})^3$ matching the simulation, and set $n_g = N/V$ where $N = 10^5$, $\sigma_v = 300$ km s$^{-1}$ and $b = 1.33$.  We scaled the covariance by a factor equal to the number of simulation boxes averaged together. Fig.\ref{fig_covmat} displays the normalised covariance matrix we calculated for the GR simulation.

It can be seen from Fig.\ref{fig_covmat} that the velocity correlation function is highly correlated on large scales because it is driven by the same low-$k$ modes of the power spectrum which cause large-scale bulk flows. The correlations reduce on smaller scales because the velocity power becomes less significant compared to the noise in the velocities produced by non-linear velocity structure. As we do not include measurement noise in our study, the large-scale correlations are further accentuated. The presence of these large-scale correlations is directly predicted by the velocity model for bulk flows \citep[e.g.][]{2009MNRAS.392..743W} and is also seen in simulations \citep[e.g.][]{2021MNRAS.502.2087T}.

\subsection{Parameter fitting}
\label{sec:parfit}

In this section we describe our approach for fitting model parameters to the measured correlation functions from Section \ref{sec_corrmeas}, using the theoretical correlation functions evaluated in Section \ref{sec_corrmod} and covariance matrix described in Section \ref{sec_corrcov}. For General Relativity we fit for the galaxy bias ($b$) and directly for the growth rate of structure ($f$), for DGP we fit $b$ and the crossover scale $r_c$, and for $f(R)$ models we fit for $b$ and the characteristic amplitude of $f_{R0}$.  Hence, all models are characterised by 2 free parameters.

In order to extract the best-fitting parameters of a model, we minimised the $\chi^2$ statistic defined by,
\begin{equation}
\chi^2 = \sum_{i,j} C^{-1}_{ij} \, \left( \xi^{sim}_i - \xi^{thry}_i \right) \left( \xi^{sim}_j - \xi^{thry}_j \right) ,
\label{eq_chi2}
\end{equation}
where $\xi^{sim} = (\xi_{\delta\delta}, \xi_{\delta v}, \xi_{vv})$ refers to the data vector of all correlation function values estimated from the simulations, and $\xi^{thry}$ refers to the data vector with the corresponding values calculated from theory. The data vectors hold the values across the full range of fitted separation scales and for all three correlation functions.

We calculated the $\chi^2$ statistic for a range of values of the fitted parameters. The parameters that produced the minimum $\chi^2$ value were taken as the best-fitting parameters in each case, and the relative probability of different fits is given by,
\begin{equation}
    Pr\propto e^{-\frac{1}{2}\chi^2} .
\end{equation}
This probability was marginalised to find the posteriors for each parameter.  We quote errors corresponding to half the width between the points corresponding to $16\%$ and $84\%$ cumulative probability, or alternatively we quote $95\%$ confidence limits.

The theoretical correlation functions were derived under the assumption that linear effects dominate cosmic dynamics. This assumption is incorrect on small scales, owing to the development of non-linearities, such as the virialisation of galaxy clusters.  Therefore, it's important to consider the range of separations over which our models should apply.

\begin{figure}
    \centering
    \includegraphics[width=\columnwidth]{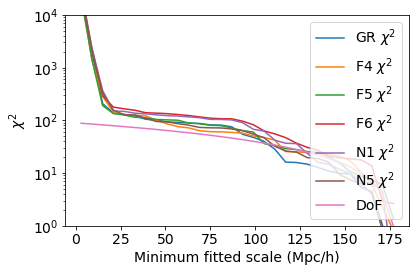}
    \caption{The $\chi^2$ value of the best-fitting model for each simulation, as a function of the minimum scale of the data considered. Also shown is the expected degrees of freedom for the fit.  When small scales ($< 20 \, h^{-1}$ Mpc) are included in the fit, the reduced $\chi^2$ value significantly increases, as the model becomes a poor fit to the data.}
    \label{fig_chimin}
\end{figure}

Fig.\ref{fig_chimin} shows the $\chi^2$ value of the best-fitting model for all our simulations, after disregarding each data point below a certain distance scale, compared to the degrees of freedom left in the data set. It can be seen that when data from small scales is included, the theoretical model is a poor fit, as the theory does not model the nonlinear dynamics of those scales. To ensure these systematic errors in the model are avoided, correlation function values below a certain separation distance were disregarded in the fitting process. After considering this analysis for all the simulations, we chose a minimum separation distance of $18 \, h^{-1}$ Mpc for all our fits (i.e., excluding the first 3 separation bins).

\section{Results}
\label{secresults}
\subsection{Fits to GR simulation}

\begin{figure}
    \centering
    \includegraphics[width=\columnwidth]{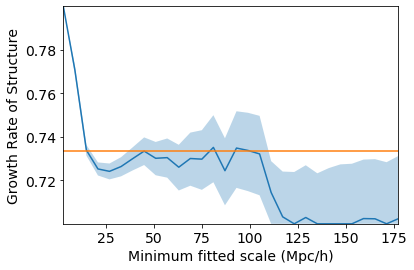}
    \caption{The best fit value and $68\%$ confidence region for the growth rate of structure parameter fit to the general relativity simulation, as a function of the minimum scale of the data considered.  The horizontal line represents the predicted value in a linear model of GR. When small scales ($< 20 \, h^{-1}$ Mpc) are included in the fit, a poor fit to the data is obtained; when the fit is restricted to the very largest scales, the statistical error increases substantially.}
    \label{fig_GRpost}
\end{figure}

The galaxy and velocity auto- and cross-correlation functions measured from the General Relativity simulation data were fit for the growth rate of structure and the galaxy bias parameter.  The growth rate of structure is not a free parameter in the $\Lambda$CDM model and is determined by the fiducial cosmology; the fiducial value of the GR simulation with $\Omega_m = 0.2814$ is $f=0.7337$ at $z=0.5$.  The $68\%$ confidence region of the marginalised growth rate is displayed in Fig.\ref{fig_GRpost} as a function of the minimum fitted scale (in the same style as the $\chi^2$ dependence shown in Fig.\ref{fig_chimin}).  For the fiducial fitting range of $18$ to $180 \, h^{-1}$ Mpc, the $68\%$ confidence range of the measured growth rate from the simulation is $f=0.7255\pm0.0030$, which is within $1\%$ of the expected value, validating the model at a precision appropriate for current and future datasets. The galaxy bias measurement for this case was $b = 1.333 \pm 0.007$.

The best-fitting model has $\chi^2=137.5$ with $79$ degrees of freedom.  Although the $\chi^2/$dof value is formally high, we note that our simulation study (using full 3D velocity information and excluding measurement noise) corresponds to an artificially accurate dataset.  Moreover, highly-correlated datasets (such as the velocity-velocity correlation function at different separations, where the correlation coefficient between adjacent bins rises to $\sim 0.99$) can be subject to some $\chi^2$ instability driven by small systematic errors in the covariance matrix (see for example the discussion in \cite{2021arXiv211013332C}).  In our case, these systematic errors could result from the approximation of an analytical Gaussian covariance matrix in linear theory.

\subsection{Fits to DGP simulations}

\begin{figure}
    \centering
    \includegraphics[width=\columnwidth]{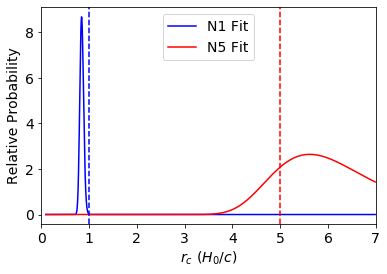}
    \caption{The posterior probability distribution of the $r_c$ parameter which characterises DGP gravity, recovered from fitting galaxy and velocity correlation function measurements from the DGP simulations, marginalising over galaxy bias. Different results are shown for simulations with fiducial values $r_c = 1c/H_0$ and $5c/H_0$, indicated by the vertical dashed lines.}
    \label{fig_DGPpost}
\end{figure}

The galaxy and velocity correlation functions of the two DGP simulations were fit for the crossover scale, $r_c$, and galaxy bias parameter, $b$, using a fitting range of $18$ to $180 \, h^{-1}$ Mpc. Fig.\ref{fig_DGPpost} shows the posterior probability distributions of the $r_c$ parameters.

The N1 simulation with fiducial $r_c = 1 \, c/H_0$ was fit with $r_c=0.852\pm0.046 \, c/H_0$ and $b=1.249\pm0.007$, with minimum $\chi^2 = 151.0$ for $79$ degrees of freedom.  The residual systematic error in our parameter fit is hence $\Delta r_c \sim 0.1 \, c/H_0$.


The N5 simulation, with a theoretical value of $r_c=5 \, c/H_0$, recovered a posterior with peak value of $r_c=5.541$. The wide distribution of this value is expected as models with high values of $r_c$ will asymptotically approach General Relativity, and will necessarily be increasingly difficult to distinguish from each other. The $r_c$ value of this distribution is higher than $4.246$ with a confidence of $95\%$. The galaxy bias was found to be $b=1.312\pm0.007$, with the $\chi^2=155.7$ with $79$ degrees of freedom. It can be seen that the galaxy bias factor found for models different from GR is slightly lower than the bias factor found for GR. This occurrence was independently noted in \cite{2019PhRvD..99f3526V} and \cite{2020JCAP...01..055V}.

\subsection{Fits to \texorpdfstring{$f(R)$}{f(R)} simulations}
\label{secfrfit}

\begin{figure}
    \centering
    \includegraphics[width=\columnwidth]{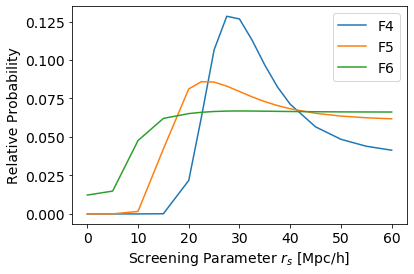}
    \caption{The relative probability of different screening scales $r_s$ to correct the $f(R)$ models using Eq.\ref{eq_fRscreen}, in a joint fit with galaxy bias. The value of $r_s = 30 \, h^{-1}$ Mpc was chosen to best account for these effects.}
    \label{fig_fRScfit}
\end{figure}

\begin{figure}
    \centering
    \includegraphics[width=\columnwidth]{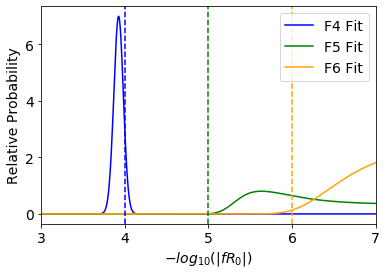}
    \caption{The posterior probability distribution of the $f_{R0}$ parameter which characterises $f(R)$ gravity, recovered from fitting galaxy and velocity correlation function measurements from the $f(R)$ simulations, marginalising over galaxy bias.  Different results are shown for simulations with $|f_{R0}| = 10^{-4}$ (F4), $10^{-5}$ (F5) and $10^{-6}$ (F6), indicated by the vertical dashed lines. F4 recovers a distinct value of the right magnitude, while F5 is loosely determined and F6 is largely indistinguishable from General Relativity.}
    \label{fig_fRpost}
\end{figure}

The correlation functions of the three $f(R)$ simulations were fit for the amplitude $f_{R0}$ and galaxy bias parameter $b$, using a fitting range of $18$ to $180 \, h^{-1}$ Mpc. As discussed in Sec.\ref{sec_corrmod}, we included the effect of screening using the phenomenological relation of Eq.\ref{eq_fRscreen}.  In order to set a fiducial value for the screening scale parameter $r_s$, we first performed a joint fit with the galaxy bias, the results of which are shown in Fig.\ref{fig_fRScfit} as the posterior probability distribution for $r_s$, marginalising over $b$. A best fitting value of $r_s=30 \, h^{-1}$ Mpc was chosen to best account for significant non-linear screening effects.

The posterior probability distributions of the $f_{R0}$ values are displayed in Fig.\ref{fig_fRpost}.  The F4 simulation could be detected as distinct from General Relativity and had a recovered parameter of $\log_{10}|f_{R0}|=-3.927\pm0.057$, and $b=1.205\pm0.007$. The fit had minimum $\chi^2=109.1$ with $79$ degrees of freedom.

The F5 simulation has a more uncertain posterior, being closer to General Relativity. There is $95\%$ confidence that the $\log_{10}|f_{R0}|$ value is below $-5.29$ with a peak value of $-5.66$, in mild tension with the fiducial value. This fit has $\chi^2=134.7$ with $79$ degrees of freedom.

The F6 simulation is largely indistinguishable from General Relativity. Its fit posterior shows that there is $95\%$ confidence that $\log_{10}(|f_{R0}|)<-6.118$. The fit has a $\chi^2=178.4$ with $79$ degrees of freedom.

\subsection{Cross-model fitting}

\begin{table}
    \centering
    \begin{tabular}{c|c|c|c}
    \hline
         Model & GR & $f(R)$ & DGP \\
    \hline
         GR&137.5&145.4&178.8\\
         F4&158.1&141.7&156.6\\
         F5&140.8&134.1&139.9\\
         F6&173.0&178.8&209.1\\
         N1&151.2&420.3&151.0\\
         N5&156.4&190.6&155.7\\
    \hline
    \end{tabular}
    \caption{Best-fitting $\chi^2$ values obtained when fitting the galaxy and velocity correlation functions of all 6 simulations (vertical) with the 3 different gravity models analysed (horizontal). We used a fitting range of $18$ to $180 \, h^{-1}$ Mpc with $79$ degrees of freedom and also varied the galaxy bias parameters.}
    \label{tab_crosschi}
\end{table}

Our model fits described above assumed that we knew in advance which modified gravity scenario described each dataset. As this would be unknown for the real Universe, we also considered whether or not the goodness of fit is sufficient to distinguish between these different models, given each simulation.  In order to test this, we fitted every simulation using all the different theoretical models, where the model that best fits the data should have the lowest $\chi^2$ statistic. The minimum $\chi^2$ statistics were calculated for each fit, and are displayed in Table \ref{tab_crosschi}.  We note that all models have two fitted parameters.

For each data set, the underlying model used in their simulation produced the best fit, except for F6 which is the most inherently indistinguishable model from GR. This implies that the different cosmological models more divergent than F6 can be differentiated by this method, although we note that the DGP simulations can be well-represented by a GR model with a different growth rate.

\subsection{Growth index fits}

\begin{figure}
    \centering
    \includegraphics[width=\columnwidth]{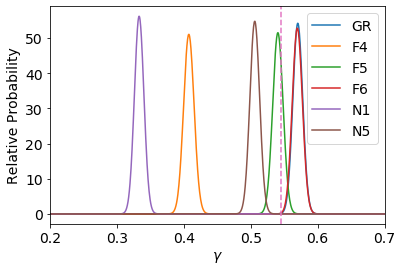}
    \caption{The posterior distribution for the growth index $\gamma$ found from fitting the correlation functions of all six cosmological models and marginalising over galaxy bias. The vertical line represents the fiducial value expected from GR.}
    \label{fig_gpost}
\end{figure}

For some models of gravity, the evolution of the growth rate of structure can be closely approximated by the relation \citep{2005PhRvD..72d3529L, 2007APh....28..481L}:
\begin{equation}
    f(z) \approx \Omega_m(z)^\gamma ,
    \label{eq_gamma}
\end{equation}
with $\gamma$ representing a free parameter that can be used to fit different models. This model-independent parameterisation has been popular in attempting to phenomenologically describe and detect modified gravity theories, so we considered how effectively this parameterisation describes our simulations.

Fig.\ref{fig_gpost} shows the posterior probability distributions of the $\gamma$ values for fits to the data from all the different simulations. The N1 model is fit with a distinct $\gamma$ value of $0.334\pm0.007$, and the F4 model produces a value of $\gamma=0.406\pm0.008$, while the other models obtain values more closely distributed to the GR model, with a value found of $0.571\pm0.007$. The standard GR value is close to $\gamma=0.545$.  The lower values of $\gamma$ found in the N1 and F4 scenarios reflect the higher growth rate relative to $\Lambda$CDM displayed in Fig.\ \ref{fig_fmodels}.

We note that the goodness-of-fit of these $\gamma$ models differs significantly between the simulations. The $\chi^2$ values of the best-fitting $\gamma$ models is the same as the growth rate fits presented in Table \ref{tab_crosschi}. Hence, the $\gamma$ parameterisation may not necessarily be convenient for characterising gravity theories that diverge from GR in a scale-dependent manner.

\section{Forecast for future surveys}
\label{secforecast}

In this section we investigate the potential of future galaxy and peculiar velocity surveys to distinguish between the different modified gravity scenarios considered in this paper.  When analysing observational data, two aspects of our framework must change. Firstly, peculiar velocities are measured using the redshift of galaxies with an independent distance measurement. This only provides the line of sight component of the peculiar velocity, as opposed to the 3-dimensional velocities considered in our simulation analysis.  Secondly, peculiar velocity measurements contain errors which grow in proportion to distance. Thus real observations of galaxies will provide less constraining information than our simulation.

To forecast the modified gravity constraints we can obtain from future surveys, we calculated a Fisher matrix.  This allows us to determine the uncertainty we expect in the model parameters we extract from certain observations, given statistical models for the data and covariance. We can then use this technique to forecast the accuracy of our analysis technique when applied to any astronomical survey, current or future, assuming any given cosmology.

\subsection{Fisher matrix theory}

We include in our forecast information arising from a joint analysis of the galaxy and peculiar velocity fields, including redshift-space distortions.  We assume a hemispherical survey of area $20{,}000$ deg$^2$ containing galaxy and peculiar velocity samples with a fiducial number density $n_g = n_v = 10^{-3} \, h^3 {\rm Mpc}^{-3}$ characteristic of surveys, and evaluate distances using a flat $\Lambda$CDM fiducial cosmology with $\Omega_{\rm m} = 0.3$. We calculated forecasts comparing two different fractional errors in distance for peculiar velocity measurements: $\epsilon = 0.05$, representing measurements from supernova surveys \citep{2017ApJ...847..128H}, and $\epsilon = 0.2$, representing measurements from fundamental plane and Tully-Fisher methods \citep{2008AJ....135.1738M,2014MNRAS.445.2677S} such as is used in 6dFGS and DESI.

The Fisher matrix calculation is divided into redshift slices and across Fourier space $(k,\mu)$, summing over $0<z<z_{\rm max}$ in bins of $\Delta z = 0.01$, $0 < k < k_{\rm max} \, h$ Mpc$^{-1}$ in bins of $\Delta k = 0.01 h$ Mpc$^{-1}$, and $0 < \mu < 1$ in bins of $\Delta \mu = 0.1$. We evaluate:
\begin{equation}
F_{ij} = \sum_{k,\mu,z} N_{\rm modes} \sum_{a,b}C_{ab}^{-1}\frac{\partial P_a}{\partial p_i}\frac{\partial P_b}{\partial p_j} ,
\label{eq_fish}
\end{equation}
where $N_{\rm modes} = V_{\rm bin}(z) \times 4\pi k^2 \, dk \, d\mu$ is the number of independent modes for each redshift bin in each Fourier cell around $(k,\mu)$, and $p_i$ are the model parameters which can be taken as $(\log_{10}(f_{R0}),b)$ or $(r_c,b)$, depending on the model being tested.  $P_a$ is a concatenation of $(P_{\delta\delta}, P_{\delta v}, P_{vv})$ for an individual Fourier mode $(k,\mu)$, computed as follows:
\begin{equation}
P_{\delta \delta}(k,\mu)=(b+f\mu)^2 P_m(k) ,
\label{eq_mpsgg}
\end{equation}
\begin{equation}
P_{\delta v}(k,\mu)=aHf\mu(b+f\mu) P_m(k) ,
\label{eq_mpsgv}
\end{equation}
\begin{equation}
P_{vv}(k,\mu)=a^2H^2f^2\mu^2 P_m(k) ,
\label{eq_mpsvv}
\end{equation}
where $P_m(k)$ is the matter power spectrum amplitude and $v$ now refers to the line-of-sight velocity.  Our covariance matrix for each mode is obtained using Eq.\ref{eqpggcov}-\ref{eqpgvcov}, where $\sigma_v = 100 \, \epsilon \, D(z)$, in terms of the distance to the redshift $D(z)$.

With the mock survey characteristics established, the only remaining inputs are those describing the history of cosmic evolution found in Eq.\ref{eq_mpsgg}-\ref{eq_mpsvv}. These can be obtained for any of the models we are considering from the numerical approach in Section \ref{sectheory}.The fiducial growth rate for the forecast was derived from the theoretical models assuming the same cosmology as in Section \ref{secsim}, and the fiducial linear galaxy bias was taken to be $b = 1$, the representative value for $L^*$ galaxies \citep{2001MNRAS.328...64N}. The exact fiducial bias value will not significantly affect results, as $b$ is marginalised in the analysis.  The derivative of $P_a$ with respect to the different model parameters can then be calculated for the range of $(k,\mu,z)$ being summed over, and substituted into Eq.\ref{eq_fish}. The value $k_{\rm max} = 0.1 \, h$ Mpc$^{-1}$ was used as an estimate of the range of scales which can be successfully modelled at low redshift \citep[e.g.,][]{2014JCAP...05..003O}.  However, we also calculated the Fisher matrix for a more conservative choice $k_{\rm max} = 0.06 \, h$ Mpc$^{-1}$, given that some studies \cite[e.g.,][]{2012MNRAS.425.2128J} show that the linear approximation for $f(R)$ can break down on larger scales. The predicted error in the measurements of a parameter is given by the square root of the corresponding diagonal component of the inverse Fisher matrix.

We note that our parameter set only includes the modified gravity variable and the galaxy bias, which provides a first  indication of the relative accuracy with which these parameters may be constrained assuming the other cosmological parameters and the matter power spectrum amplitude are known.  However, we note that a full analysis would vary the values of all cosmological parameters.

\subsection{Fisher matrix results}

\begin{figure}
    \centering
    \includegraphics[width=\columnwidth]{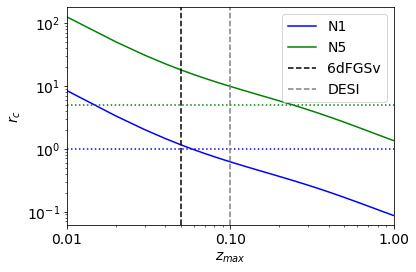}
    \caption{The forecast error of a combined galaxy and peculiar velocity survey to determine the cross-over scale $r_c$ of a DGP universe, as a function of the maximum survey redshift. For the purposes of this plot, a fractional distance error $\epsilon = 0.2$ was used for measurements. We consider two different fiducial values $r_c = (1c/H_0, 5c/H_0)$, indicated by the horizontal dotted lines. Scales up to $k_{\rm max} = 0.1 \, h$ Mpc$^{-1}$ were summed over. When the forecast error is one third of the fiducial values, we consider that the surveys will be sensitive enough to detect the gravity model of its universe with 3-$\sigma$ significance. The vertical dotted lines show indicative maximum redshifts of the 6dFGSv and DESI PV surveys.}
    \label{fig_DGPfish}
\end{figure}

\begin{figure}
    \centering
    \includegraphics[width=\columnwidth]{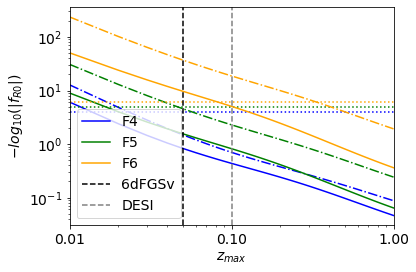}
    \caption{The forecast error of a combined galaxy and peculiar velocity survey to determine the characteristic parameter $f_{R0}$ of an $f(R)$ universe, as a function of the maximum survey redshift. A $\epsilon=0.2$ fractional error of the distance was used for measurements. Solid lines represent a forecast using scales up to $k_{\rm max} = 0.1 \, h$ Mpc$^{-1}$, while dot-dash lines use scales up to $k_{\rm max} = 0.06 \, h$ Mpc$^{-1}$ for a more conservative analysis.  We consider three different fiducial values $f_{R0} = (10^{-4}, 10^{-5}, 10^{-6})$, indicated by the horizontal dotted lines. The vertical dotted lines show indicative maximum redshifts of the 6dFGS and DESI PV surveys.}
    \label{fig_fRfish}
\end{figure}

Figs.\ref{fig_DGPfish} and \ref{fig_fRfish} show the forecast error in the characteristic parameters of the DGP and $f(R)$ models, with respect to the maximum redshift of the galaxies being measured in the survey. We consider a series of models with the same fiducial values as adopted in our simulations, which we label (F4,F5,F6) and (N1,N5) as before. These fiducial values are represented by the horizontal lines in the figures.

At the point that the predicted error falls below a third of the fiducial value, we consider that the expected characteristic parameter should be detected with a $3\sigma$ confidence level. If no such value is detected at that point, we can be equally confident that the assumed cosmic model is ruled out at this significance level.

Fig.\ref{fig_DGPfish} and Fig.\ref{fig_fRfish} show that if the cosmic models of F4 and F5 were descriptions of the universe we live in, peculiar velocities from existing surveys may already be able to identify their effects. The less divergent simulations, N1, N5, and F6, haven't met the $3\sigma$ threshold using current data and so are likely currently undistinguished from the $\Lambda$CDM model using this test. The vertical lines in Figs.\ref{fig_DGPfish} and \ref{fig_fRfish} represent the effective maximum redshifts of the 6dFGS velocity survey and the DESI peculiar velocity survey, although some details of these surveys may vary slightly from the parameters used in the calculation of the error.

\begin{table*}
    \centering
    \begin{tabular}{c|c|c|c|c|c|c}
        \hline
        & & & $k_{\rm max} = 0.1 \, h \, {\rm Mpc}^{-1}$ & & & \\
        \hline
        & & $\epsilon=0.05$ & & & $\epsilon=0.2$ & \\
        & $z_{max}$ & $\sigma$ (z=0.1) & $\sigma$ (z=0.5) & $z_{max}$ & $\sigma$ (z=0.1) & $\sigma$ (z=0.5) \\
        \hline
        F4 $(|{\rm log}|f_{R0}||)$ & 0.027 & 0.233 & 0.049 & 0.032 & 0.435 & 0.096 \\
        F5 $(|{\rm log}|f_{R0}||)$ & 0.031 & 0.405 & 0.104 & 0.047 & 0.819 & 0.142 \\
        F6 $(|{\rm log}|f_{R0}||)$ & 0.124 & 2.458 & 0.672 & 0.238 & 5.054 & 0.810 \\
        N1 $(r_cH_0/c)$ & 0.097 & 0.323 & 0.069 & 0.222 & 0.618 & 0.168 \\
        N5 $(r_cH_0/c)$ & 0.288 & 4.912 & 1.109 & 0.802 & 9.793 & 2.635 \\
        \hline
    \end{tabular}
    \caption{Fisher matrix forecasts for the constraining power of future peculiar velocity surveys on modified gravity, assuming the 5 fiducial models considered in this study, for maximum wavenumber $k_{\rm max} = 0.1 \, h$ Mpc$^{-1}$.  The 2nd column lists the maximum redshift ($z_{max}$) that a survey would require in order to achieve a likely detection of the cosmological model for each of the modified gravity theories. We then list the errors in the characteristic parameters, given a survey that reaches to maximum redshifts of $z=0.1$ and $z=0.5$. These numbers were found for cases where the fractional error in the distance measurement was $\epsilon=0.05$ and $\epsilon=0.2$, characteristic of supernova and fundamental plane/Tully-Fisher relation measurements, respectively.}
    \label{tab_forcast1}
\end{table*}

\begin{table*}
    \centering
    \begin{tabular}{c|c|c|c|c|c|c}
        \hline
        & & & $k_{\rm max} = 0.06 \, h \, {\rm Mpc}^{-1}$ & & & \\
        \hline
        & & $\epsilon=0.05$ & & & $\epsilon=0.2$ & \\
        & $z_{max}$ & $\sigma$ (z=0.1) & $\sigma$ (z=0.5) & $z_{max}$ & $\sigma$ (z=0.1) & $\sigma$ (z=0.5) \\
        \hline
        F4 $(|{\rm log}|f_{R0}||)$ & 0.046 & 0.462 & 0.078 & 0.055 & 0.700 & 0.169 \\
        F5 $(|{\rm log}|f_{R0}||)$ & 0.081 & 1.292 & 0.255 & 0.141 & 2.283 & 0.498 \\
        F6 $(|{\rm log}|f_{R0}||)$ & 0.529 & 10.255 & 2.093 & 0.963 & 18.713 & 4.006 \\
        N1 $(r_cH_0/c)$ & 0.157 & 0.606 & 0.097 & 0.328 & 0.874 & 0.247 \\
        N5 $(r_cH_0/c)$ & 0.448 & 9.063 & 1.520 & 1.526 & 13.600 & 3.927 \\
        \hline
    \end{tabular}
    \caption{Fisher matrix forecasts displayed in the same style as Table \ref{tab_forcast1}, for maximum wavenumber $k_{\rm max} = 0.06 \, h$ Mpc$^{-1}$.}
    \label{tab_forcast6}
\end{table*}

Table \ref{tab_forcast1} shows detection threshold forecasts from the Fisher matrix with $k_{\rm max} = 0.1 \, h$ Mpc$^{-1}$, giving the maximum redshifts needed to obtain 3-$\sigma$ detections of $f_{R0}$ and $r_c$ for different fiducial choices of those parameters. Table \ref{tab_forcast6} provides the same information for a more conservative choice $k_{\rm max} = 0.06 \, h$ Mpc$^{-1}$, using information from scales more confident to conform to linear predictions. The other columns of these tables display the expected accuracy of detecting a model for redshift surveys that reach to $z=0.1$ (characteristic of DESI) and $z=0.5$ (characteristic of a future SNe survey such as that described by \cite{2017ApJ...847..128H}), for fractional distance error expected from using supernova distances and galaxy scaling relations. For galaxy scaling errors, only F4 and F5 have a detection threshold below $z_{max}=0.1$ , but F6 and N1 are detectable in a survey that extends to $z=0.5$.

Interpolating these results we find that for the upcoming generation of surveys such as DESI, with $\epsilon=0.2$, $z_{max}=0.1$ and $k_{\rm max} = 0.1 \, h$ Mpc$^{-1}$, $f_{R0}$ values are constrained to less than $10^{-5.5}$, and $r_c$ to greater than $0.4c/H_0$. A future survey with $\epsilon=0.2$ and $z_{max}=0.5$ would constrain $f_{R0}$ to less than $10^{-6.5}$, and $r_c$ to greater than $2.9c/H_0$. For an analysis only using scales $k < 0.06 \, h$ Mpc$^{-1}$, we find $f_{R0}<10^{-4.79}$ for DESI and $10^{-5.66}$ for future surveys. These will be competitive constraints on these parameters.

The simulations analysed in Section \ref{secresults} can't be directly compared to these forecasts, since the forecasts take into account redshift-space distortion effects, distance-dependent errors, and only utilise the line of sight velocity component, which are not included in our simulations. Though our simulations provide a proof-of-concept for our analysis and theory, our forecasts are more likely to reflect results expected from real surveys.

\section{Conclusion}
\label{secconc}

Dark Energy remains a theory of cosmic expansion that currently doesn't have a strong theoretical basis.  In hopes of finding a deeper understanding of gravity, various modified gravity theories have been proposed that could also explain cosmic expansion. To develop techniques to distinguish which of these theories can describe our universe, we have used cosmological simulations generated in popular modified gravity scenarios to investigate the growth rate of cosmic structure as measured by the distribution of direct peculiar velocities induced by the different models of gravity.

In our analysis technique we numerically calculated the linear approximation of the galaxy auto-correlation, the galaxy-velocity cross-correlation, and the velocity auto-correlation for three different gravity models: General Relativity, $f(R)$, and DGP. For $f(R)$ we took into account the dependence of the growth rate on scale. We then fit these functions to the correlation functions extracted from modified gravity simulations.

We find that linear-theory models, with an additional phenomenological screening term in case of $f(R)$ gravity, are able to describe our simulations with acceptable accuracy on scales $>20 \, h^{-1}$ Mpc, and can successfully distinguish the fiducial parameters of simulations generated in popular modified gravity scenarios with a small residual systematic error. The linear theory model with screening that we employ was found to be sufficient for our purposes, but a possible future improvement to this method would be to use emulators for the density and velocity power-spectra computed from simulations \citep{2021PhRvD.103l3525R, 2021arXiv210904984A}. Likewise we could also consider more sophisticated perturbation theory models such as Lagrangian perturbation theory for the correlation function \citep{2020JCAP...01..055V} and effective field theory-inspired extensions for the power spectrum \citep{2021JCAP...04..039A}.

We can test the ability of this analysis to differentiate different theories by cross-fitting the simulations with alternate theoretical models. As can be seen in Table \ref{tab_crosschi}, the $\chi^2$ statistic shows that the underlying models of simulations more divergent than the F6 model could be differentiated by this analysis. This analysis also showed that the approximate growth index $\gamma$ was not necessarily a good model-independent fitting parameter, as it can result in poor fits owing to not capturing the correct amplitude or scale-dependence of the growth rate of structure in alternative scenarios.

We also presented Fisher matrix forecasts, showing that this method can place competitive limits on possible models with current surveys. Upcoming peculiar velocity surveys, such as DESI, 4MOST and future supernova surveys, will place accurate constraints that will start to compete with other cosmological probes. A future PV survey to $z=0.5$ with $20\%$ fractional distance errors is forecast to constrain values to $f_{R0}<10^{-6.45}$ for $f(R)$ gravity and $r_c>2.88c/H_0$ for nDGP.

In summary, we have shown that the auto- and cross-correlation functions of peculiar velocities and galaxies can be used to describe and distinguish between cosmic simulations of different scenarios of gravitational physics. These results, and the Fisher matrix forecasts we calculated for future surveys, indicate that this analysis method can be a competitive probe for investigating the physics of the universe.

\section*{Acknowledgements}

We thank the anonymous referee for several helpful suggestions which clarified and improved the study. This project received financial support through an Australian Government Research Training Program Scholarship awarded to SL, and a Swinburne University Postgraduate Research Award awarded to RT.  RR was supported by the Australian Government through Australian Research Council Discovery Project DP160102705 and Laureate Fellowship Project FL180100168.

We thank Baojiu Li for making the ELEPHANT simulations available, from which our modified gravity simulations are drawn.

\section*{Data Availability}

The data underlying this article will be shared on reasonable request to the corresponding author.

\bibliographystyle{mnras}
\bibliography{example}

\begin{appendix}

\section{Analytical Covariance}
\label{secapp}

In this section we derive expressions for the analytical covariance of the galaxy and velocity correlation functions.  We start by determining the covariance of the galaxy auto-correlation function.  The angle-averaged galaxy auto-correlation function $\xi_{\delta\delta}$ in a narrow shell around separation $r$ can be expressed in terms of the galaxy auto-power spectrum $P_{\delta\delta}(\vec{k})$ as,
\begin{equation}
  \xi_{\delta\delta}(r) = \int \frac{d^3\vec{k}}{(2\pi)^3} P_{\delta\delta}(\vec{k}) \, j_0(kr) .
\end{equation}
The covariance of this correlation function at different separations $r$ and $r'$ then follows as,
\begin{equation}
\begin{split}
  & {\rm Cov} \left[ \xi_{\delta\delta}(r), \xi_{\delta\delta}(r') \right] = \\ & \int \frac{d^3\vec{k}}{(2\pi)^3} \int \frac{d^3\vec{k}'}{(2\pi)^3} {\rm Cov} \left[ P_{\delta\delta}(\vec{k}) , P_{\delta\delta}(\vec{k}') \right] \, j_0(kr) \, j_0(k'r') .
\end{split}
\end{equation}
We can simplify this expression by assuming that power spectrum modes for different wavevectors are uncorrelated inside the survey box of volume $V$, such that only modes with $\vec{k}' = \pm \vec{k}$ have a non-zero covariance.  Integrating $\vec{k}'$ over the resulting delta-functions and assuming the power spectrum is isotropic, we find,
\begin{equation}
\begin{split}
  & {\rm Cov} \left[ \xi_{\delta\delta}(r), \xi_{\delta\delta}(r') \right] = \\ & \frac{2}{V} \int \frac{dk \, k^2}{2\pi^2} {\rm Cov} \left[ P_{\delta\delta}(k), P_{\delta\delta}(k) \right] \, j_0(kr) \, j_0(kr') ,
\end{split}
\end{equation}
where the covariance of the galaxy auto-power spectrum fluctuations for each mode is given by Eq.\ref{eqpggcov}.  For a broad separation bin $i$ in the range $r_1 < r < r_2$, we average the correlation function over $r$ with volume weighting, following Eq.\ref{eqxiave}.  In this case,
\begin{equation}
  {\rm Cov} \left[ \overline{\xi}^i_{\delta\delta} , \overline{\xi}^j_{\delta\delta} \right] = \frac{2}{V} \int \frac{dk \, k^2}{2\pi^2} {\rm Cov} \left[ P_{\delta\delta}(k) , P_{\delta\delta}(k) \right] \, W^i_{\delta} \, W^j_{\delta} ,
\end{equation}
where
\begin{equation}
W^i_{\delta\delta} = \frac{\int_{r_1}^{r_2} dr \, r^2 \, j_0(kr)}{\int_{r_1}^{r_2} dr \, r^2} = \frac{3 \left[ r_2^2 \, j_1(kr_2) - r_1^2 \, j_1(kr_1) \right]}{k \left( r_2^3 - r_1^3 \right)}
\end{equation}
is evaluated for each separation bin.

Repeating this derivation for the velocity auto-correlation, we find for a narrow separation bin:
\begin{equation}
\begin{split}
  & {\rm Cov} \left[ \xi_{vv}(r), \xi_{vv}(r') \right] = \frac{2}{V} \int \frac{dk \, k^2}{2\pi^2} {\rm Cov} \left[ P_{vv}(k) , P_{vv}(k) \right] \\ & \left[ j_0(kr) - \frac{2 j_1(kr)}{kr} \right] \left[ j_0(kr') - \frac{2 j_1(kr')}{kr'} \right] ,
\end{split}
\end{equation}
in terms of the velocity auto-power spectrum $P_{vv}(k)$.  For a broad separation bin,
\begin{equation}
  {\rm Cov} \left[ \overline{\xi}^i_{vv} , \overline{\xi}^j_{vv} \right] = \frac{2}{V} \int \frac{dk \, k^2}{2\pi^2} {\rm Cov} \left[ P_{vv}(k) , P_{vv}(k) \right] \, W^i_{vv} \, W^j_{vv} ,
\end{equation}
where
\begin{equation}
W^i_{vv} = \frac{\int_{r_1}^{r_2} dr \, r^2 \, \left[ j_0(kr) - \frac{2 j_1(kr)}{kr} \right]}{\int_{r_1}^{r_2} dr \, r^2} = \frac{3 \int_{k r_1}^{k r_2} du \, u^2 \left[ j_0(u) - \frac{2 j_1(u)}{u} \right]}{k^3 \left( r_2^3 - r_1^3 \right)} ,
\end{equation}
and the covariance of the velocity auto-power spectrum fluctuations is given by Eq.\ref{eqpvvcov}.

Finally for the galaxy-velocity cross-correlation, we find for a narrow separation bin:
\begin{equation}
\begin{split}
  & {\rm Cov} \left[ \xi_{\delta v}(r), \xi_{\delta v}(r') \right] = \\ & \frac{2}{V} \int \frac{dk \, k^2}{2\pi^2} {\rm Cov} \left[ P_{\delta v}(k) , P_{\delta v}(k) \right] \, j_1(kr) \, j_1(kr') ,
\end{split}
\end{equation}
in terms of the galaxy-velocity cross-power spectrum $P_{gv}(k)$.  For a broad separation bin,
\begin{equation}
  {\rm Cov} \left[ \overline{\xi}^i_{\delta v} , \overline{\xi}^j_{\delta v} \right] = \frac{2}{V} \int \frac{dk \, k^2}{2\pi^2} {\rm Cov} \left[ P_{\delta v}(k) , P_{\delta v}(k) \right] \, W^i_{\delta v} \, W^j_{\delta v} ,
\end{equation}
where
\begin{equation}
W^i_{\delta v} = \frac{\int_{r_1}^{r_2} dr \, r^2 \, j_1(kr)}{\int_{r_1}^{r_2} dr \, r^2} = \frac{3 \int_{k r_1}^{k r_2} du \, u^2 \, j_1(u)}{k^3 \left( r_2^3 - r_1^3 \right)} ,
\end{equation}
where the variance of the galaxy-velocity cross-power spectrum fluctuations for each mode is given by Eq.\ref{eqpgvcov}.

\bsp	
\label{lastpage}

\end{appendix}

\end{document}